 \documentclass[journal]{IEEEtran}
\ifCLASSINFOpdf
\else
\fi
\usepackage[cmex10]{amsmath}

\usepackage{xcolor}
\usepackage{tikz}
\usepackage{booktabs}
\usetikzlibrary{intersections,calc}
\usetikzlibrary{calc}
\usetikzlibrary{arrows}
\usepackage{pgfplots}
\usepackage{algorithmic}
\usepackage{algorithm}
\usepackage{cite}

\newcommand*{\blue}{\textcolor{black}}
\DeclareRobustCommand*{\ora}{\overrightarrow}
\begin{document}
%
\title{Efficient Creation of Datasets for Data-Driven \\ Power System Applications}

\author{Andreas~Venzke, 
        Daniel~K.~Molzahn, 
        and~Spyros~Chatzivasileiadis \vspace{-0.3cm}
\thanks{A. Venzke and S. Chatzivasileiadis are with the Department of Electrical Engineering, Technical University of Denmark, 2800 Kgs. Lyngby, Denmark e-mail: \{andven, spchatz\}@elektro.dtu.dk.}
\thanks{D. K. Molzahn is with the School of Electrical and Computer Engineering, Georgia Institute of Technology, Atlanta, GA 30313, USA as well as the Energy Systems Division, Argonne National Laboratory, Lemont, IL 60439 USA, e-mail: molzahn@gatech.edu.}
\thanks{This work  is  supported  by  the  multiDC  project,  funded  by  Innovation  Fund Denmark, Grant Agreement No. 6154-00020B.}
}




\maketitle

\begin{abstract}
Advances in data-driven methods have sparked renewed interest for applications in power systems. Creating datasets for successful application of these methods has proven to be very challenging, especially when considering power system security. This paper proposes a computationally efficient method to create datasets of secure and insecure operating points. We propose an infeasibility certificate based on separating hyperplanes that can a-priori characterize large parts of the input space as insecure, thus significantly reducing both computation time and problem size. Our method can handle an order of magnitude more control variables and creates balanced datasets of secure and insecure operating points, which is essential for data-driven applications. While we focus on N-1 security and uncertainty, our method can extend to dynamic security. For PGLib-OPF networks up to 500 buses and up to 125 control variables, we demonstrate drastic reductions in unclassified input space volumes and computation time, create balanced datasets, and evaluate an illustrative data-driven application.
\end{abstract} 




\section{Introduction}
Recent advances in data-driven methods have shown substantial potential for power system applications including security assessment under uncertainty~\cite{wehenkel2012automatic,donnot2018fast,Cremer2019TPWRS,Halilbasic_PSCC_2018,arteaga2019deep,duchesne2018using}, e.g., by rapidly estimating line flows~\cite{donnot2018fast}, training accurate security classifiers~\cite{Cremer2019TPWRS}, and applying these classifiers in the context of data-driven security-constrained optimal power flow \cite{Halilbasic_PSCC_2018} and deep learning toolboxes \cite{arteaga2019deep}. The performance of these methods, however, relies on the quality of the underlying dataset. As historical data is often limited and does not contain many abnormal situations, the datasets have to be enriched through simulation. This, however, is a highly computationally demanding task. The resulting datasets should be balanced between secure and insecure samples to improve classifier performance, take into consideration all degrees of freedom of the system, and be able to accurately represent the security boundary. In this work, we propose an efficient method to create datasets with these properties for data-driven applications in power systems.

The steady-state operational constraints are described by the AC optimal power flow (AC-OPF) problem. The degrees of freedom of the system, i.e., the inputs characterizing each operating point, are defined by the control variables, which in the AC-OPF problem are generator active power and voltage set-points. By defining these, the remaining state variables are determined by solving the AC power flow equations~\cite{molzahn2018fnt}. Even for medium-sized systems, the resulting number of control variables renders the task of creating datasets covering a wide range of operating points very computationally challenging.

To address this challenge, we can directly classify operating points that are infeasible with respect to the AC-OPF problem as insecure and avoid any further stability or static security assessment. Ref.~\cite{molzahn2017computing} formulated infeasibility certificates with respect to the AC-OPF problem that are based on hyperspheres which certify a wide range of operating points a-priori as insecure. Inspired by \cite{molzahn2017computing}, our previous work in \cite{Thams2019} used such certificates to generate large datasets, reducing the input space and decreasing computation time, while considering both \mbox{N-1} security and small-signal stability. Both works \cite{molzahn2017computing} and \cite{Thams2019} consider systems with up to 11 control variables. Instead of hyperspheres, this paper proposes the use of separating hyperplanes, which, among other important benefits, allows us to consider numbers of control variables that are at least an order of magnitude greater than previous methods (up to 125 in our test cases). 

Another popular approach to create such datasets is through importance sampling, e.g.,~\cite{Krishnan2011, hamon2016importance}. In power systems, however, the initial sampling space is largely unbalanced, i.e., the volume of insecure space is several orders of magnitude larger than the secure space, and, as we observed in \cite{Thams2019}, it can be challenging to obtain an adequate number of secure samples. In this work, we show how our proposed method can lead to a balanced dataset, as it enables us to sample from inside the secure space. A related strand of research uses historical data that is enriched through sampling methods such as composite modelling approaches and vine-copulas~\cite{Sun2016, Konstantelos2019TPWRS}. However, this can neglect parts of the secure space or might not capture abnormal operating regions. 

To create representative datasets for data-driven power system applications, we propose a computationally efficient method which a) can deal with high input dimensionality (our test cases have up to 125 control variables), b) provides a detailed description of the security boundary, and c) creates \emph{balanced} classes.  We apply this method to AC-OPF problems including N-1 security and uncertainty in power injections. The main contributions of our work are:
\begin{enumerate}
    \item We propose an infeasibility certificate based on separating hyperplanes. This certificate is computed using convex relaxations of AC-OPF problems and considers both N-1 security and uncertainty. Compared to the hypersphere-based method proposed in \cite{molzahn2017computing}, our algorithm shows two key improvements: First, separating hyperplanes allow the classification of substantially larger parts of the input space as insecure. Second, as these hyperplanes form a convex polytope, efficient methods to sample uniformly from inside the remaining unclassified space are available. Based on these, we propose an efficient algorithm to maximize the volume of the input space classified a-priori as insecure. 
 \item We evaluate this algorithm on PGLib-OPF networks with up to 500 buses and numbers of control variables up to 125. Compared to initial normalized input space volumes of $1$ (i.e.,\,$10^0$) based on specified control variable bounds, the infeasibility certificates reduce the unclassified input space volumes significantly, \blue{with reduced volumes ranging from $10^{-2}$ up to $10^{-40}$}.   
    \item We propose a computationally efficient method to create datasets for data-driven power system applications which can handle systems where the number of control variables is at least one order of magnitude greater than state-of-the-art methods (e.g.,  \cite{Thams2019}). Computing infeasibility certificates allows us to efficiently characterize the security boundary in detail and sample from inside the secure space. We create balanced datasets for PGLib-OPF networks up to 500 buses and train neural network classifiers as an illustrative data-driven application.
\end{enumerate}
This paper is structured as follows: In Section~\ref{Sec:ACOPF}, we describe the AC-OPF problem including N-1 security and uncertainty, and its convex relaxation. In Section~\ref{Sec:Meth}, we outline our proposed methodology to create datasets, including the infeasibility certificate, boundary description, and sampling from inside the secure space. Section~\ref{Sec:Sim} presents simulation results on  PGLib-OPF networks up to 500 buses. Section~\ref{Sec:Con} concludes.

\section{Optimal Power Flow Formulation} \label{Sec:ACOPF}
This section presents the AC-OPF formulations necessary for deriving the dataset creation methodology. In particular, we formulate the N-1 security-constrained preventive AC-OPF problem considering uncertainty in power injections, and its quadratic convex (QC) relaxation. For a detailed survey on AC-OPF and convex relaxations of the AC-OPF, the reader is referred to \cite{cain2012history,molzahn2018fnt}. Here, for brevity, we build our formulation upon the AC-OPF formulation of \cite{coffrin2016qc} to facilitate the derivation of the QC relaxation. We use the QC relaxation as it represents a good trade-off between computational complexity and tightness of the relaxation \cite{coffrin2016qc}. Note that the following derivations could be readily extended via the many other convex relaxations of the power flow equations~\cite{molzahn2018fnt}.

\subsection{Security-Constrained AC-OPF under Uncertainty}
\label{sec:formulation}
A power system is defined by its set $\mathcal{N}$ of buses. A subset of those buses, which are denoted by $\mathcal{G}$, have a controllable generator connected. A second subset denoted by $\mathcal{U}$, which can be either generation or load buses, are subject to uncertain power injections. It is assumed that all buses of the power system are connected by a set $(i,j) \in \mathcal{L}$ of power lines from bus $i$ to bus $j$. To ensure the N-1 security criterion during operation, we consider the potential outage of a list of critical candidate lines defined by the set $\mathcal{C} \subset \mathcal{L}$. Note that we define the first entry of $\mathcal{C}$ to correspond to the intact system state $\{0\}$, i.e., no transmission line is outaged. \blue{The term $\mathcal{L}^c$ denotes the set of intact power lines for outage  $c \in \mathcal{C}$. For the intact system state, the set $\mathcal{L}^0$ corresponds to the set $\mathcal{L}$.}

The optimization variables in the security-constrained AC-OPF are the complex bus voltages $V_k^c$ for each bus $k \in \mathcal{N}$ and contingency $c \in \mathcal{C}$, the complex power dispatch of generator $S_{G_k}^c$ for each bus $k \in \mathcal{G}$ and contingency $c \in \mathcal{C}$, and the uncertain complex power injections $S_{U_k}$ for each bus $k \in \mathcal{U}$. The uncertain power injections do not change upon outage of system components, i.e., $S_{U} = S_{U}^c, \, \, \forall c \in \mathcal{C}$. We assume that the uncertain reactive power injection $Q_{U} = \Im \{S_U\}$ is determined through a fixed power factor $\cos \phi$ in relation to the uncertain active power injection $P_{U} = \Re \{S_U\}$, i.e., $Q_{U} = \sqrt{\tfrac{1-\cos^2 \phi}{\cos^2 \phi}} \, P_{U}$. \blue{If the power factor is not constant, then the reactive power injection $Q_{U}$ can be modelled as a separate variable, i.e., as a separate degree of freedom in the dataset creation method.} The following constraints must be satisfied for the intact system and for each contingency $c \in \mathcal{C}$:
\begin{subequations}
\label{opf}
\begin{align}
  &  (V_k^{\text{min}})^2 \leq V_k^c (V_k^c)^* \leq (V_k^{\text{max}})^2 &   \forall k \in \mathcal{N}  \label{Vmax} \\
  &  S_{G_k}^{\text{min}} \leq S_{G_k}^c \leq S_{G_k}^{\text{max}} &   \forall k \in \mathcal{G}  \label{Smax} \\
  &  |S_{ij}^c| \leq S_{ij}^{\text{max}} & \forall (i,j) \in \mathcal{L}^{c} \label{Sijmax} \\
  &  S_{G_k}^c - S_{D_k} + S_{U_k} = \sum_{(k,j) \in \mathcal{L}^{c}} S_{kj}^c &  \forall k \in \mathcal{N} \label{Sbal} \\
  & S_{ij}^c = (Y_{ij}^c)^* V_i^c (V_i^c)^* - (Y_{ij}^c)^*V_i^c (V_j^c)^* &  \forall (i,j) \in \mathcal{L}^{c} \label{Sij} \\
 &     S_{U_k}^{\text{min}} \leq S_{U_k} \leq S_{U_k}^{\text{max}} &   \forall k \in \mathcal{U}  \label{SUmax} \\
  & \theta^{\text{min}}_{ij} \leq \angle (V_i^c (V_j^c)^*) \leq \theta^{\text{max}}_{ij} &  \forall (i,j) \in \mathcal{L}^{c}  \label{angle} 
\end{align}
\end{subequations}
The bus voltage magnitudes are constrained in \eqref{Vmax} by upper and lower limits $V_k^{\text{min}}$ and $V_k^{\text{max}}$. The superscript $*$ denotes the complex conjugate. Similarly, the generators' complex power outputs are limited in \eqref{Smax} by upper and lower bounds $S_{G_k}^{\text{min}}$ and $S_{G_k}^{\text{max}}$. The inequality constraints for complex variables are defined as bounds on the real and imaginary parts. The apparent power flow $S_{ij}$ on the line from $i$ to $j$ is upper bounded in \eqref{Sijmax} by $S_{ij}^{\text{max}}$. The nodal complex power balance \eqref{Sbal} including the load $S_D$, generation $S_G$ and uncertain injections $S_U$ has to hold for each bus. The apparent power flow $S_{ij}$ on the line from $i$ to $j$ is defined in \eqref{Sij}. The term $Y$ denotes the admittance matrix of the power grid. Constraint \eqref{SUmax} models minimum and maximum bounds $S_{U_k}^{\text{min}}$, $S_{U_k}^{\text{max}}$ on the uncertain injections. \blue{The flow on the line from $i$ to $j$ is limited in \eqref{angle} by a lower and upper limit on angle differences $\theta^{\text{min}}_{ij}$ and $\theta^{\text{max}}_{ij}$, respectively. Please note that for most instances the following holds: $\theta^{\text{min}}_{ij} = -\theta^{\text{max}}_{ij}$.} 

We consider preventive actions in the security-constrained AC-OPF formulation, i.e., the generator set-points remain fixed during an outage. As a result, we include the following linking constraints between the intact system state and the outaged system states:
\begin{subequations}
\label{link_outage}
\begin{align}
    & |V_k^0| = |V_k^c| &  \quad \forall k \in \mathcal{G}, \; \forall c \in \mathcal{C}\backslash \{0\} \label{VG_link} \\
    & P_{G_k}^0 =  P_{G_k}^c &  \quad \forall k \in \mathcal{G}\backslash \{\text{slack}\},  \; \forall c \in \mathcal{C}\backslash \{0\} \label{PG_link} 
\end{align}
\end{subequations}
The active power dispatch is denoted as $P_{G}$, i.e., $P_G =  \Re \{S_G\}$. The first constraint sets the generator voltage set-points $|V_k|$ of the outaged system states to the values from the intact system state. The second constraint does the same for the active power generation dispatch, excluding the slack bus which compensates the difference in active power losses.
\subsection{Quadratic Convex (QC) Relaxation}
The QC relaxation proposed in \cite{coffrin2016qc} uses convex envelopes of the polar representation of the AC-OPF problem to relax the dependencies among voltage variables. As proposed in \cite{Lavaei2012, coffrin2016qc}, an additional auxiliary matrix variable $W^c$ is introduced for the intact system state and each contingency $c \in \mathcal{C}$, which denotes the product of the complex bus voltages: 
\begin{align}
    W_{ij}^c = V_i^c (V_j^c)^* & \quad \forall c \in \mathcal{C} \label{Wnonconvex}
\end{align}
This allows reformulation of \eqref{Vmax}, \eqref{Sij}, \eqref{angle}, and \eqref{VG_link} as:
\begin{subequations}
\label{Wconstraints}
\begin{align}
      &  (V_k^{\text{min}})^2 \leq W_{kk}^c \leq (V_k^{\text{max}})^2  &  \forall k \in \mathcal{N} \label{Wmax} \\
       & S_{ij}^c = (Y_{ij}^c)^* W_{ii}^c - (Y_{ij}^c)^* W_{ij}^c  & \forall (i,j) \in \mathcal{L}^{c} \label{WSij} \\
           & S_{ji}^c = (Y_{ij}^c)^* W_{jj}^c- (Y_{ij}^c)^* (W_{ij}^c)^* \hspace*{-30pt}  & \forall (i,j) \in \mathcal{L}^{c} \label{WSji} \\
             & \tan(\theta^{\text{min}}_{ij}) \leq \tfrac{\Re \{W_{ij}^c\}}{\Im \{W_{ij}^c\}} \leq \tan(\theta^{\text{max}}_{ij}) \hspace*{-30pt} & \forall (i,j) \in \mathcal{L}^{c} \label{Wangle} \\
               &  W_{kk}^0 = W_{kk}^c  & \forall k \in \mathcal{G}, \; \forall c \in \mathcal{C}\backslash \{0\} \label{VG_link2} 
\end{align}
\end{subequations}
The non-convexity is encapsulated in the voltage product \eqref{Wnonconvex}. To obtain a convex relaxation, the non-convex constraint \eqref{Wnonconvex} is removed from the optimization problem and variables for voltages, $v_i^c \angle \theta_i^c \, \,\forall i \in \mathcal{N} \, \, \forall c \in \mathcal{C}$, and squared current flows, $l_{ij}^c \, \,\forall (i,j) \in \mathcal{L}^{c} \, \,  \forall c \in \mathcal{C}$, are added. The following convex constraints and envelopes are introduced for the intact system state and each contingency $c \in \mathcal{C}$~\cite{coffrin2016qc}:
\begin{subequations}
\small
\label{QC}
\begin{align}
  & W_{kk}^c = \left\langle v_k^2 \right\rangle^T  &  \forall k \in \mathcal{N} \label{QC1}\\
  & \Re \{W_{ij}^c\} =   \left\langle \left\langle v_i^c v_j^c \right\rangle^M   \left\langle \cos (\theta_i^c - \theta_j^c) \right\rangle^C \right\rangle^M  &  \forall (i,j) \in \mathcal{L}^{c}\\
  & \Im \{W_{ij}^c\} =   \left\langle \left\langle v_i^c v_j^c \right\rangle^M   \left\langle \sin (\theta_i^c - \theta_j^c) \right\rangle^S \right\rangle^M &   \forall (i,j) \in \mathcal{L}^{c}\\
  & S_{ij}^c + S_{ji}^c = Z_{ij}^c l_{ij}^c &  \forall (i,j) \in \mathcal{L}^{c}\\
  & |S_{ij}^c|^2 \leq W_{ii}^c l_{ij}^c &  \forall (i,j) \in \mathcal{L}^{c} \label{QCend}
\end{align}
\end{subequations}
The superscripts $T,M,C,S$ denote convex envelopes for the square, bilinear product, cosine, and sine functions, respectively. The term $Z_{ij}$ denotes the line impedance. Refer to \cite{coffrin2016qc} for the complete QC formulation. The resulting relaxation of the preventive security-constrained AC-OPF under uncertainty is a second-order cone program (SOCP) that minimizes an objective function, e.g., generation cost, subject to \eqref{Smax}--\eqref{Sbal}, \eqref{SUmax}, \eqref{PG_link}, \eqref{Wconstraints}, and \eqref{QC}. 

\section{Methodology to Create Datasets} \label{Sec:Meth}
The goal of the following methodology is to create a dataset which maps operating points described by the input vector~$x$ to a power system security classification, e.g., secure or insecure. The dataset should be balanced between secure and insecure samples, take into consideration the degrees of freedom of the system, and have a detailed description of the security boundary. The power system security classification we consider is feasibility with respect to the N-1 security-constrained AC-OPF problem under uncertainty defined in \eqref{opf} and \eqref{link_outage}. The resulting dataset can be complemented with further assessment of dynamic security criteria, e.g. small-signal stability \cite{Thams2019}. 
 The input vector $x$, i.e., the control variables that define the relevant degrees of freedom, is defined as follows:
\begin{align}
  & x = \begin{bmatrix}
               P_{G_i}^0\\
               |V_j^0|\\
               P_{U_k}
            \end{bmatrix} & \forall i \in \mathcal{G}\backslash \{\text{slack}\},\; \forall j \in \mathcal{G},\; \forall k \in \mathcal{U} \label{x_def}
\end{align}
 Using the input $x$, all other states in the AC-OPF problem can be determined by solving the non-linear AC power flow equations. The minimum and maximum bounds on input vector $x^{\text{max}}$ and $x^{\text{min}}$ are defined in \eqref{Vmax}, \eqref{Smax}, and~\eqref{SUmax}. \blue{Please note that in the formulation of the QC relaxation, we use the variable $v_j^0$ instead of the eliminated variable $|V_j^0|$.}

The main challenge in creating a representative and balanced dataset is the large number of control variables. The dimensionality of the input vector $x$ grows substantially with increasing system size. For instance, the IEEE 118-bus system has 72 control variables, i.e., the dimensionality $|x|$ is 72. A na\"ive approach to create a dataset would be to sample with a prespecified discretization interval, e.g., by specifying 10 steps in each dimension of the control variables, $x_1, x_2, x_3, \ldots$. For the 118-bus system, this would require power flow solutions for $10^{72}$ operating points, which is computationally intractable. Further, as we will empirically show in Section~\ref{Sec:volume}, large parts of the input space $x \in [x^{\text{min}}, x^{\text{max}}]$ are infeasible. As a result, identifying secure samples by na\"ively sampling from the entire input space is not possible for larger test cases.

To address these challenges, we present an efficient method for creating such datasets. First, to a-priori classify large parts of the input space as insecure, we propose an infeasibility certificate based on separating hyperplanes in Section~\ref{sec:infeas}. Focusing on the unclassified regions, we then characterize the security boundary in detail in Section~\ref{sec:boundary}. Finally, we sample inside the secure space in Section~\ref{sec:inside}.

\subsection{Constructing Infeasibility Certificates}\label{sec:infeas}
We propose an infeasibility certificate which can a-priori certify regions in which the non-convex security-constrained AC-OPF problem under uncertainty is infeasible. This exploits the following property of a convex relaxation: if a relaxation is infeasible for a given operating point, the original non-convex problem is also guaranteed to be infeasible for that operating point. The proposed infeasibility certificate has three components: First, we employ bound tightening to tighten both the QC relaxation and the input bounds; this better approximates the secure region, while also reducing the sample space. Second, we propose an infeasibility certificate based on separating hyperplanes. Third, we present an efficient algorithm to maximize the input region classified as infeasible.

\subsubsection{Bound Tightening Algorithms} \label{sec:BT} The tightness of the QC relaxation relies on the tightness of the envelopes used in \eqref{QC} including the envelopes on cosine and sine terms. These in turn depend on the tightness of the bounds on the voltage magnitudes and angle differences. The goal of bound tightening is to iteratively tighten voltage magnitudes and angle differences, and, as a result, obtain a tighter relaxation. In the context of our work, the benefits of bound tightening are twofold: First, it tightens the QC relaxation, i.e., shrinks its feasible space, making the infeasibility certificate based on separating hyperplanes more effective, and second, it allows us to directly tighten the bounds on the input vector $x$.

We use two bound tightening algorithms from the literature: First, we rely on a computationally lightweight bound tightening technique for the branch angle differences $\theta_{ij}^{\text{min}}$ and $\theta_{ij}^{\text{max}}$ in \eqref{angle} from \cite{Shchetinin2019}. Second, we use an optimization-based bound tightening algorithm from \cite{sundar2018optimization} which tightens the voltage magnitude bounds at each bus $V^{\text{max}}, V^{\text{max}}$ in \eqref{Vmax}, and further tightens the angle differences for each line $\theta_{ij}^{\text{min}}$ and $\theta_{ij}^{\text{max}}$ in \eqref{angle}. For this purpose, we iteratively solve convex optimization problems to calculate the maximum and minimum values that the optimization variable under study, i.e., a voltage magnitude or a voltage angle difference, can obtain in the relaxed problem. Note that by tightening one variable bound, it may be possible to further tighten a previously tightened bound. This procedure can be executed for a defined number of iterations or until a fixed point is reached. As a final step in the bound tightening, we compute the tightened bounds for the input vector $x$, i.e., the bounds on active power of generators and uncertain injections. All inputs $x$ which are outside the tightened minimum and maximum input bounds $x^{\text{BT,min}}$, $x^{\text{BT,max}}$ are guaranteed to be infeasible with respect to the non-convex AC-OPF problem. We calculate the volume of the remaining unclassified input space volume $V_{BT}$, normalized by the originally specified bounds on $x$:
\begin{align}
   V_{BT} = \prod_{k \in \mathcal{X}}  \tfrac{x_k^{\text{BT,max}} - x_k^{\text{BT,min}}}{x_k^{\text{max}} - x_k^{\text{min}}} \label{eq:V_BT}  
\end{align}
The input set $\mathcal{X}$ is defined as $\mathcal{X}: \{ \mathcal{G}\backslash \{\text{slack}\}, \, \mathcal{G}, \, \mathcal{U} \}$. 

\subsubsection{Separating Hyperplanes}  \label{sec:HP}
\begin{figure}
    \centering
    \begin{tikzpicture}


\node (o) at (0, 0) {};
\draw [gray!20,fill=gray!20] (4.88,2.2) ellipse (2 and 0.6);
\draw [green,fill=green] plot [smooth cycle,fill=green] coordinates {(3.2,2) (4.2,2.2) (4.85,2.7)
 (5.3,2.3) (6.6,2.2) (5.3,1.9) (4.85,1.7) (4.2,1.9)};
\draw[name path=circle,fill=red,fill opacity=0.2, draw = red, thick] (2.3, 1.3) circle (0.99); 
\draw[draw=black,thick] (o) rectangle ++ (7,3);
\draw[line width=0.75pt,black,->,black,thick] (2.3, 1.3)  -- (3, 2);
\draw[line width=3pt,yellow,thick](3, 2) -- (2.006,3);
\draw[line width=3pt,yellow,thick] (3, 2) -- (4.96,0);
\fill [yellow,fill opacity=0.2] (0,0) -- (0,3) -- (2.006,3) -- (4.96,0);
\node (P1label) at (2.1, 1.1) {$\hat{x}$};
\node (P1) at (2.3, 1.3) {$\mathbf{+}$};
\node (P2) at (3, 2) {$\mathbf{+}$};
\node (P2label) at (2.75, 2) {$x^*$};
\node (P2label) at (3.4, 1.4) {$|\overrightarrow{n}|=R^*$};
\end{tikzpicture}
         \begin{tikzpicture} 
    \begin{axis}[%
    hide axis,
    xmin=0,
    xmax=0.1,
    ymin=0,
    ymax=0.1,
    legend style={draw=white!15!black,legend cell align=left},
    legend columns=3,
    legend style={font=\footnotesize},
    ]
    \addlegendimage{mark options={solid, red, thick}, mark=o,only marks,mark size=4pt}
    \addlegendentry{Hypersphere};
        \addlegendimage{yellow, line width = 3pt}
    \addlegendentry{Hyperplane \,};
        \addlegendimage{mark options={solid, black, thick}, mark=square,only marks,mark size=4pt}
    \addlegendentry{Input space $x \in [x^{\text{min}}\,x^{\text{max}}]$};
       \addlegendimage{mark options={solid, gray!20, thick}, mark=o,only marks,mark size=4pt}
    \addlegendentry{\makebox[0pt][l]{Feasible space (QC relaxation) \, \,}}

    \addlegendimage{empty legend}
    \addlegendentry{\,}
     \addlegendimage{mark options={solid, green, thick}, mark=star,only marks,mark size=4pt}
    \addlegendentry{\makebox[0pt][l]{Feasible space (AC-OPF)}}
    \end{axis}
\end{tikzpicture}
\vspace{-4.75cm}
    \caption{Illustrative example of the differences between the infeasibility certificates using hyperspheres and hyperplanes. For a given infeasible point $\hat{x}$, the closest point $x^*$ is computed which is feasible to the QC relaxation. The normal vector $\ora{n}$ is perpendicular to the feasible space of the QC relaxation. The feasible space of the non-convex AC-OPF problem is contained within that of the convex QC relaxation. All points inside the hypersphere or all points that are on the left side of the hyperplane are guaranteed to be infeasible with respect to both the QC relaxation and the non-convex AC-OPF problem, respectively. Note that the sets are not drawn to scale.}
    \label{Hyper}
\end{figure}
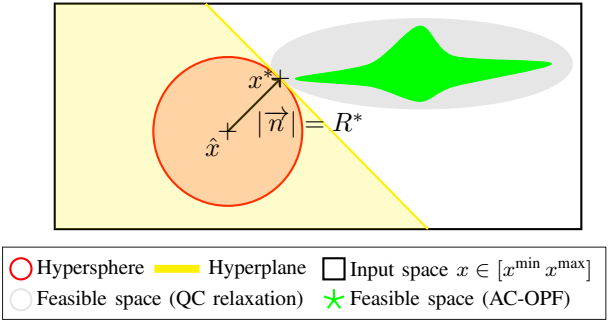
We next propose an infeasibility certificate based on separating hyperplanes. Consider a particular operating point $\hat{x}$ that is infeasible with respect to the non-convex security-constrained AC-OPF. We solve the following optimization problem to compute the closest dispatch $x^*$ which is feasible to the convex QC relaxation:
\begin{subequations}
\label{closest_feas_qc} 
\begin{align}
    \min_{x,S,S_U,S_G,v,\theta,l,W,R} \, & \;\; R \\
    \text{s.t.}\, \, &  \eqref{Smax}\text{--}\eqref{Sbal}, \eqref{SUmax}, \eqref{PG_link}, \eqref{Wconstraints}, \eqref{QC}, \eqref{x_def} \label{QCfeasibility} \\
    & \sqrt{\sum_{k \in \mathcal{X}} (x_k - \hat{x}_k)^2} \leq R \label{x_norm}
\end{align}
\end{subequations}
If the obtained radius $R^*$ is greater than zero, i.e., the operating point $\hat{x}$ is infeasible with respect to the relaxation, no operating point $x$ exists which is closer to $\hat{x}$ than the obtained point $x^*$. This property has been used in \cite{molzahn2017computing} to construct infeasibility certificates in the form of hyperspheres and ellipses by assigning different weights to the components in \eqref{x_norm}. Here, we propose to use hyperplanes as infeasibility certificates in order to significantly enlarge the volume classified as infeasible: \newline
\textit{Proposition 1:} For a given infeasible point $\hat{x}$, if the solution to \eqref{closest_feas_qc} yields a non-zero radius $R^*$ and optimal solution $x^*$, all vectors $x$ which fulfill the following criterion are infeasible with respect to the AC-OPF constraints \eqref{opf} and \eqref{link_outage}:
\begin{align}
    \overrightarrow{n}^T (x - x^*) < 0 \label{hyperplane_def}
\end{align}
The normal of the hyperplane is defined as $ \overrightarrow{n}: = x^* - \hat{x}$ and the operator $T$ denotes the transpose.\newline
\textit{Proof of Proposition 1:} Proof by contradiction: Assume there exists a feasible point $\tilde{x}$ that is inside the region classified as infeasible by the hyperplane: $\overrightarrow{n}^T (\tilde{x} - x^*) < 0$. As the feasible space of optimization problem \eqref{closest_feas_qc} is convex, it must hold that any linear combination between $\tilde{x}$ and $ x^*$ is also feasible: $\lambda \tilde{x} + (1-\lambda) x^*, \lambda \in [0,1] $. Then, there exists a point $\tilde{x}^* = \lambda^* \tilde{x} + (1-\lambda^*) x^* $ which has a radius $\tilde{R}^*$ to the initial infeasible point $\hat{x}$ that is smaller than $R^*$. Since the optimization problem \eqref{closest_feas_qc} is convex, we obtained the globally optimal solution $x^*$ with the smallest radius $R^*$. As a result, there cannot exist an input $\tilde{x}$ that has a smaller radius than $R^*$. We have shown by contradiction that there cannot exist a feasible point $\tilde{x}$ that is inside the region classified as infeasible by the hyperplane. The infeasibility certificate with respect to the non-convex AC-OPF problem \eqref{opf} and \eqref{link_outage} follows from the property that infeasibility with respect to the QC relaxation constraints \eqref{QCfeasibility} is sufficient to ensure infeasibility with respect to \eqref{opf} and \eqref{link_outage}. \blue{Alternatively, we can show that Proposition 1 is true by taking the first-order Taylor expansion of constraint \eqref{x_norm} at the optimal solution $x^*$. For convex sets, first-order Taylor expansions of nonlinear constraints are always separating hyperplanes \cite{boyd2004convex}.} 

An illustrative comparison of both infeasibility certificates is shown in Fig.~\ref{Hyper}. By solving the same optimization problem, it is evident that the infeasibility certificate based on hyperplanes is able to classify significantly larger spaces as infeasible. This is quantitatively analysed through simulation studies in Section~\ref{Sec:Sim_Comp}.

\subsubsection{An Efficient Algorithm to Minimize the Unclassified Input Space}
\begin{algorithm}[t]
\caption{Computing Infeasibility Certificates}
\begin{algorithmic}[1]
\STATE Run bound tightening and obtain $x^{\text{BT,min}}$ and $x^{\text{BT,max}}$
\STATE Set iteration count: $k \leftarrow 0$  
\STATE Initialize unclassified region $A^{(0)} x \leq b^{(0)}$: \\
\quad $A^{(0)}:= [\mathbf{I}^{|x| \times |x|} \, \,  -\mathbf{I}^{|x| \times |x|}]^T$ \\
\quad $b^{(0)}: \, \,= [(x^{\text{BT,max}})^T \, \,(x^{\text{BT,min}})^T]^T$
\WHILE{$k \leq N_1$}
\STATE draw random $x^{(k)}$ from inside  $A^{(k)} x \leq b^{(k)}$
\STATE solve \eqref{closest_feas_qc} with $\hat{x}:=x^{(k)}$ and obtain $x^*$
\IF{$R > 0$}
\STATE reduce unclassified region by adding hyperplane: \\ \quad $A^{(k+1)} = [(A^{(k)})^T \, \, \overrightarrow{n}]^T$ \\ \quad $b^{(k+1)} \, \,= [(b^{(k)})^T \, \, \overrightarrow{n}^T x^*]^T$
\ENDIF
\STATE $k \leftarrow k+1$
\ENDWHILE.
\end{algorithmic}
\label{ALG_IF}
\end{algorithm}
Using the infeasibility certificate, we propose an efficient algorithm to maximize the portion of the input space that can be classified a-priori as infeasible. Our algorithm relies on an insight related to the hyperplanes: together with the initial input space restriction, subsequent hyperplanes form a convex polytope which can be described as $A x \leq b$. We can write the row of $A$ and entry in $b$ corresponding to the hyperplane in \eqref{hyperplane_def} as $A_k:= \overrightarrow{n}^T$ and $b_k:= \overrightarrow{n}^T x^*$. Efficient methods to sample uniformly from inside a convex polytope are available, e.g., ``Hit-and-Run'' sampling \cite{kroese2013handbook}. This allows us to iteratively construct hyperplanes while sampling only inside the currently unclassified region. Thus, the hyperplane certificates facilitate a significant improvement on the ``rejection'' sampling approach used with hypersphere certificates in~\cite{molzahn2017computing,Thams2019}.

The steps of the algorithm to compute infeasibility certificates are detailed in Algorithm~\ref{ALG_IF}. We start with a description of the convex polytope restricted to the tightened input bounds. We iteratively sample uniformly from inside the convex polytope and add identified hyperplanes until we reach an upper iteration limit of $N_1$ samples. This ensures that only samples which have not yet been classified as infeasible by previously added hyperplanes are considered in optimization problem~\eqref{closest_feas_qc}. In Section~\ref{Sec:volume}, we will demonstrate the performance of this algorithm on a range of PGLib-OPF networks up to 500 buses by calculating the remaining unclassified volume as the volume of the convex polytope $A^{(N_1)} x \leq b^{(N_1)}$. This shows substantial reductions of unclassified input space volumes. \blue{An alternative approach to the proposed algorithm using separating hyperplanes could be to directly construct a linear outer approximation of the convex feasibility set defined by \eqref{QCfeasibility}. This could be achieved by applying e.g. ``Hit-and-Run'' sampling to \eqref{QCfeasibility}. This is subject of our future work.}

\subsection{Security Boundary Identification}\label{sec:boundary}
After computing the infeasibility certificates, we perform sampling and directed walks, similar to \cite{Thams2019}, to obtain a detailed description of the security boundary. For this purpose, we first uniformly draw a large number $N_2$ of samples from the convex polytope describing the remaining unclassified input region: $A^{(N_1)} x \leq b^{(N_1)}$. For each sample, we first run AC power flows for the intact and the outaged system states and check if any of the constraints in \eqref{opf} are violated. If not, we add the current point to the dataset as a feasible point, otherwise as an infeasible point. If constraints are violated, we run additional AC power flows for which we enforce the reactive power limits of generators, i.e., if any generator violates its reactive power limit it is converted from a $PV$ to a $PQ$ bus in the power flow. This is based on the observation that reactive power limits are often the only constraints violated. If the obtained power flow solutions satisfy all constraints in \eqref{opf}, \blue{the voltage-adjusted point is added to the dataset as feasible sample, i.e., the voltage set-points of generators in $x$ are updated accordingly.} If both stages are not feasible, we solve the following non-convex optimization problem which computes the closest feasible dispatch to the non-convex AC-OPF problem in \eqref{opf} and \eqref{link_outage}:
    \begin{subequations}
\label{closest_feas_ac} 
\begin{align}
    \min_{x,V,S_U,S_G,S,R} \, & \;\; R \\
    \text{s.t.}\, \, &  \eqref{opf},\eqref{link_outage}, \eqref{x_def}, \eqref{x_norm} 
\end{align}
\end{subequations}
 We add the obtained locally optimal point $x^*$ to the dataset as feasible point. We repeat this procedure for all $N_2$ samples and obtain as a result a detailed security boundary description. 

\subsection{Sampling from Inside the Secure Space}\label{sec:inside}
To obtain a more detailed description of the entire secure space, we fit a multivariate normal distribution $\mathcal{N}$ to the feasible points obtained. For this purpose, we estimate both the mean $\mu$ and the covariance matrix $\Sigma$ from the feasible data points. To bias the sampling towards inside the boundary, we reduce the magnitude of all entries of the covariance matrix, i.e., $\Sigma_{\text{red}} = s_{\text{red}} \cdot \Sigma$, by a constant scaling factor $s_{\text{red}} < 1$. We draw a large number, denoted $N_3$, of samples from $\mathcal{N}(\mu,\Sigma_{\text{red}})$. For each of these samples, we first run AC power flows for the intact and the outaged system states, check feasibility with respect to all AC-OPF constraints, and add the sample with the corresponding classification to the dataset. If the sample is infeasible, we run a second round of AC power flows in which we enforce the generators' reactive limits and again evaluate the feasibility with respect to all AC-OPF constraints. If the sample is feasible, we add it to the dataset with the generator voltage set-points adjusted accordingly. Our simulations indicate that sampling from a multivariate normal distribution $\mathcal{N}(\mu,\Sigma_{\text{red}})$ results in identification of feasible samples inside the secure space. We did not observe improvements by fitting a Gaussian mixture model.

\section{Simulation and Results}  \label{Sec:Sim}
We analyse the performance of our proposed methodology for a range of test cases from the PGLib-OPF networks. First, we compare the proposed infeasibility certificate based on separating hyperplanes with the certificate based on hyperspheres from \cite{molzahn2017computing}. Second, we compute the volume of the unclassified input space using the infeasibility certificates and show substantial reductions. Third, we create balanced datasets and demonstrate their applicability using an illustrative data-driven application.

\subsection{Simulation Setup}
In the following, we first evaluate our proposed methods on 13 PGLib-OPF networks (v19.05)~\cite{PGLIB} up to 500 buses for which we do not consider N-1 security and uncertainty, i.e., we use the test cases as specified in \cite{PGLIB}. Second, we use two test cases for which we include both N-1 security and uncertainty. We use \textit{case39\_epri} and \textit{case162\_ieee\_dtc} with the following line contingencies $\mathcal{C}=\{0, 7, 22, 24, 36, 43\}$ and $\mathcal{C}=\{\blue{0}, 6, 8, 24, 50, 128\}$, respectively. \blue{These lines correspond to the following bus pairs $\{$--, 3-18, 12-13, 14-15, 22-23, 26-28$\}$ and $\{$--, 2-7, 3-14, 8-13, 16-17, 50-125$\}$, respectively.} We assume the same parameters for the outaged system state as for the intact system state. Furthermore, we place three wind farms with rated power of 500 MW and consider three uncertain loads with $\pm50$\% variability, i.e., a total of six uncertain power injections, at buses $\mathcal{U}=\{3, 21, 27, 4, 25, 28\}$ for \textit{case39\_epri} and $\mathcal{U}=\{60, 90, 145, 3, 8, 52\}$  for \textit{case162\_ieee\_dtc}. For all uncertain injections, we assume a power factor $\cos \phi=1$. 

Note that all inputs $x$ are normalized with respect to their maximum and minimum limits as specified in \cite{PGLIB}, i.e., if $x$ has dimension $|x|$, then $x \in [0, 1]^{|x|}$. This normalization step is standard practice for many data-driven applications including neural networks and improves performance \cite{tensorflow2015-whitepaper}. 

For both AC power flow and AC optimal power flow computations, we rely on M{\sc atpower} \cite{zimmerman2010matpower} with the I{\sc popt} solver for AC-OPF problems \cite{wachter2006implementation}. For the bound tightening, we use the implementations in \cite{Shchetinin2019, sundar2018optimization}. \blue{Note that we adapted the implementations in both \cite{Shchetinin2019} and \cite{sundar2018optimization} to include uncertainty in power injections by modeling them as generators with active power limits corresponding to the defined uncertainty set and no reactive power capability, i.e., the lower and upper reactive power limits are set to zero (as the power factor is assumed to be 1, $\cos \phi=1$).} We only tighten the bounds of the intact system state, \blue{i.e., the bounds of the outaged system states are not tightened,} and we run the optimization-based bound tightening for up to three iterations. Extension of these toolboxes to the full N-1 case is a direction for future work. We use MOSEK~\cite{mosek} to solve the QC relaxation. 

To approximate the volumes of the convex polytopes describing the remaining unclassified input space, we use a volume approximation toolbox in C++ \cite{emiris2018practical} \blue{which handles floating point precision issues}. Note that an exact volume computation is considered intractable for dimensionality 10 or higher \cite{emiris2018practical}. The relative approximation error threshold is set to be less than one order of magnitude, which is sufficiently accurate for our purposes since we compute volumes of spaces several orders of magnitudes smaller than the initial volume. 
\subsection{Comparison of Infeasiblity Certificates} \label{Sec:Sim_Comp}
\begin{figure}
    \centering
    \begin{footnotesize}
%
%
\definecolor{mycolor1}{rgb}{0.00000,0.44700,0.74100}%
\definecolor{mycolor2}{rgb}{0.85000,0.32500,0.09800}%
\definecolor{mycolor3}{rgb}{0.92900,0.69400,0.12500}%
\definecolor{mycolor4}{rgb}{0.49400,0.18400,0.55600}%
\begin{tikzpicture}

\begin{axis}[%
width=3.5cm,
height=2.25cm,
scale only axis,
xmin=0,
xmax=50,
xlabel style={font=\color{white!15!black}},
xlabel={Iterations},
ymode=log,
ymin=1e-05,
ytick={1e-05,1e-04,1e-03,1e-02,1e-01,1},
xtick={0, 10, 20, 30, 40, 50},
ymax=1,
yminorticks=true,
ylabel style={font=\color{white!15!black}},
ylabel={Unclass. Volume},
axis background/.style={fill=white},
legend pos=outer north east,
legend cell align={left},
grid=minor,
]

\addplot[const plot, color=red,dashed, line width = 1pt] table[row sep=crcr] {%
0	1\\
1	0.3575\\
2	0.3511\\
3	0.3494\\
4	0.3494\\
5	0.296\\
6	0.2206\\
7	0.2206\\
8	0.2206\\
9	0.2182\\
10	0.2182\\
11	0.2182\\
12	0.2182\\
13	0.2182\\
14	0.2163\\
15	0.2163\\
16	0.2163\\
17	0.2163\\
18	0.2163\\
19	0.2163\\
20	0.2163\\
21	0.2163\\
22	0.2163\\
23	0.2163\\
24	0.2163\\
25	0.2163\\
26	0.2163\\
27	0.2163\\
28	0.2078\\
29	0.2078\\
30	0.2078\\
31	0.2078\\
32	0.2078\\
33	0.2071\\
34	0.2071\\
35	0.2071\\
36	0.2071\\
37	0.2071\\
38	0.1684\\
39	0.1684\\
40	0.1684\\
41	0.1684\\
42	0.1684\\
43	0.1684\\
44	0.1684\\
45	0.1682\\
46	0.1682\\
47	0.1682\\
48	0.1682\\
49	0.1682\\
50	0.1203\\
51	0.1203\\
52	0.1203\\
53	0.1065\\
54	0.1065\\
55	0.1065\\
56	0.1065\\
57	0.1003\\
58	0.1003\\
59	0.1003\\
60	0.1003\\
61	0.1003\\
62	0.1003\\
63	0.1003\\
64	0.1003\\
65	0.1003\\
66	0.1003\\
67	0.1003\\
68	0.1003\\
69	0.1003\\
70	0.1003\\
71	0.1003\\
72	0.1003\\
73	0.1003\\
74	0.1003\\
75	0.1003\\
76	0.1003\\
77	0.1003\\
78	0.0919\\
79	0.0919\\
80	0.0919\\
81	0.0919\\
82	0.0919\\
83	0.0919\\
84	0.0919\\
85	0.0919\\
86	0.0919\\
87	0.0919\\
88	0.0919\\
89	0.0919\\
90	0.0917\\
91	0.0917\\
92	0.0917\\
93	0.0917\\
94	0.0917\\
95	0.0917\\
96	0.0917\\
97	0.0917\\
98	0.0917\\
99	0.0917\\
100	0.0917\\
};
\addlegendentry{Hyperspheres (39 bus)}

\addplot[const plot, color=red, line width = 1pt] table[row sep=crcr] {%
0	1\\
1	0.000192123906794935\\
2	0.000150940355601456\\
3	3.09551326638371e-05\\
4	2.70101660017262e-05\\
28	2.28533851876767e-05\\
57	2.24937246293035e-05\\
100	2.24937246293035e-05\\
};
\addlegendentry{Hyperplanes (39 bus)}

\addplot[const plot, color=blue,dashed, line width = 1pt] table[row sep=crcr] {%
0	1\\
1	0.9225\\
2	0.9178\\
3	0.793\\
4	0.71\\
5	0.7079\\
6	0.7079\\
7	0.7042\\
8	0.7036\\
9	0.6969\\
10	0.6646\\
11	0.6646\\
12	0.5059\\
13	0.4635\\
14	0.4635\\
15	0.2921\\
16	0.2907\\
17	0.2841\\
18	0.2841\\
19	0.2841\\
20	0.2841\\
21	0.2841\\
22	0.2841\\
23	0.2841\\
24	0.2841\\
25	0.2841\\
26	0.2841\\
27	0.2834\\
28	0.2834\\
29	0.2827\\
30	0.2827\\
31	0.2827\\
32	0.2827\\
33	0.2827\\
34	0.2827\\
35	0.2827\\
36	0.2819\\
37	0.2801\\
38	0.2801\\
39	0.2445\\
40	0.2445\\
41	0.2348\\
42	0.2348\\
43	0.2348\\
44	0.2348\\
45	0.2348\\
46	0.2348\\
47	0.2346\\
48	0.2346\\
49	0.2346\\
50	0.2346\\
51	0.2346\\
52	0.2346\\
53	0.2346\\
54	0.2346\\
55	0.2343\\
56	0.2343\\
57	0.2343\\
58	0.2343\\
59	0.2343\\
60	0.2268\\
61	0.2268\\
62	0.2236\\
63	0.2236\\
64	0.2236\\
65	0.2236\\
66	0.2236\\
67	0.2148\\
68	0.2144\\
69	0.2144\\
70	0.2144\\
71	0.2144\\
72	0.2144\\
73	0.2144\\
74	0.2144\\
75	0.2144\\
76	0.2144\\
77	0.2138\\
78	0.2138\\
79	0.2138\\
80	0.2138\\
81	0.2138\\
82	0.2138\\
83	0.2138\\
84	0.2115\\
85	0.2113\\
86	0.2113\\
87	0.2113\\
88	0.2113\\
89	0.2113\\
90	0.2113\\
91	0.2113\\
92	0.2108\\
93	0.2108\\
94	0.2108\\
95	0.2108\\
96	0.2108\\
97	0.2108\\
98	0.2108\\
99	0.2108\\
100	0.2108\\
};
\addlegendentry{Hyperspheres (162 bus)}

\addplot[const plot, color=blue, line width = 1pt] table[row sep=crcr] {%
0	1\\
1	0.00198416211297823\\
2	0.000345988284913061\\
3	0.000174324446290684\\
4	0.000159468236832074\\
5	4.47253916553718e-05\\
11	3.91644316516024e-05\\
12	3.42229121234063e-05\\
14	2.31720192243835e-05\\
27	2.187929793212e-05\\
28	2.11461357623304e-05\\
36	2.07736225034844e-05\\
38	2.08603384575041e-05\\
48	2.08065964859991e-05\\
49	2.06970904072197e-05\\
53	2.0390099083237e-05\\
55	2.01479655808189e-05\\
64	2.00628763154993e-05\\
69	2.00272030470286e-05\\
95	1.95730141520999e-05\\
100	1.95730141520999e-05\\
};
\addlegendentry{Hyperplanes (162 bus)}

\end{axis}

\end{tikzpicture}%
    \caption{For \textit{case39\_epri} and \textit{case162\_ieee\_dtc}, we compare the remaining unclassified volume between an infeasibility certificate based on hyperspheres from \cite{molzahn2017computing} and the proposed certificate in Section~\ref{sec:HP} based on hyperplanes. \blue{Note that these results cannot be directly compared to those of Table~\ref{Unclassified}, as here only the active generator set-points are assumed as degrees of freedom. In Table~\ref{Unclassified}, generator voltage set-points and uncertain injections are also considered.} }
    \label{Fig_Comp}
    \end{footnotesize}
    \vspace{-0.5cm} 
\end{figure}
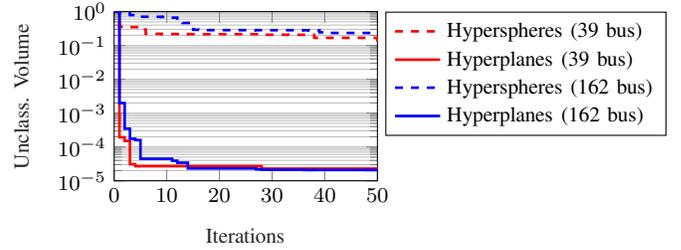
We compare the infeasibility certificate based on hyperspheres proposed in \cite{molzahn2017computing} with the infeasibility certificate based on hyperplanes proposed in Section~\ref{sec:HP}. The main metric for comparison is the volume of the remaining unclassified space after applying the infeasibility certificates. We consider \textit{case39\_epri} and \textit{case162\_ieee\_dtc}. We use the QC relaxation for both certificates with all the bounds tightened as described in Section~\ref{sec:BT}. We only consider the active power generation in the input variables $x$ and we do not consider \mbox{N-1} security or uncertainty. For the certificate based on hyperplanes, we follow the algorithm outlined in Algorithm~\ref{ALG_IF}. For the certificate based on hyperspheres, we assume that in each iteration we draw a random sample from the entire input space, and if it is infeasible we compute the closest feasible input $x$ by solving \eqref{closest_feas_qc}. If the distance is non-zero, we have obtained an infeasibility certificate. We use Monte Carlo sampling with $10^6$ samples to estimate the volume of the unclassified space not covered by the hyperspheres. 

Fig.~\ref{Fig_Comp} shows the variation of the unclassified input volume with up to 50 iterations. We make the following observations from the results. First, the hyperplane certificates shrink the unclassified region by four orders of magnitude more than the hyperspheres, i.e., the unclassified volumes evaluate to $10^{-5}$ versus $10^{-1}$ compared to the initial unit hypercube's normalized volume of $1$. Second, the algorithm using hyperplanes requires significantly fewer iterations. After the first iteration, the hyperplanes classify a substantially larger space as infeasible than the hyperspheres after 50 iterations. The reasons for this are twofold: First, as evident in Fig.~\ref{Hyper}, certificates based on separating hyperplanes cover a larger volume than hyperspheres for the same sample and second, the hyperplanes enable the use of efficient methods for sampling uniformly from inside the associated convex polytope~\mbox{\cite{kroese2013handbook, emiris2018practical}}. 
\subsection{Estimating Unclassified Volumes for PGLib-OPF Networks} \label{Sec:volume}
\begin{table}
\caption{Unclassified input volumes for PGLib-OPF networks}
\begin{tabular}{l c c c c c}
\toprule
Power system case &  $|x|$ & $V^{BT}$ & $|HP|$ & $V^{HP}$  & $\tfrac{-\log_{10}(V)}{|x|}$\\
\midrule
                           \multicolumn{6}{c}{AC-OPF without N-1 security and without uncertainty} \\
              \midrule        
               \textit{case3\_lmbd} &   4 &  6.3e-02 &  28 &  3.3e-02 & 37.0\% \\ 
                   \midrule
               \textit{case5\_pjm} &   7 &  1.0e+00 &  99 &  6.9e-03 & 30.9\% \\ 
                    \midrule
              \textit{case14\_ieee} &   6 &  2.4e-01 &  54 &  6.9e-04 & 52.7\% \\ 
                  \midrule
          \textit{case24\_ieee\_rts} &  20 &  9.2e-01 & 184 &  2.3e-06 & 28.2\% \\ 
              \midrule
              \textit{case30\_ieee} &   7 &  6.2e-03 &  61 &  8.8e-06 & 72.2\% \\ 
                  \midrule
              \textit{case39\_epri} &  19 &  9.9e-02 & 203 &  7.0e-08 & 37.7\% \\ 
                  \midrule
              \textit{case57\_ieee} &  10 &  3.8e-02 & 231 &  4.9e-06 & 53.1\% \\ 
                  \midrule
          \textit{case73\_ieee\_rts} &  62 &  1.0e+00 & 608 &  6.1e-16 & 24.5\% \\ 
              \midrule
             \textit{case118\_ieee} &  72 &  1.7e-02 & 1000 &  1.6e-17 & 23.3\% \\ 
                 \midrule
         \textit{case162\_ieee\_dtc} &  23 &  6.1e-04 & 371 &  1.5e-11 & 47.1\% \\ 
                 \midrule
            \textit{case200\_tamu} &  69 &  9.3e-01 & 1000 &  6.0e-11 & 14.8\% \\ 
                \midrule
             \textit{case300\_ieee} & 125 &  1.0e-12 & 1000 &  3.4e-40 & 31.6\% \\ 
                 \midrule
            \textit{case500\_tamu} & 111 &  8.6e-02 & 1000 &  5.4e-26 & 22.8\% \\ 
             \midrule
                           \multicolumn{6}{c}{AC-OPF considering N-1 security and uncertainty} \\
              \midrule        
              \textit{case39\_epri}  &  25  &  2.6e-01   &  271     & 2.0e-05 & 18.8\% \\ 
                      \midrule        
               \textit{case162\_ieee\_dtc } &  29  & 2.2e-04 &  394     & 6.0e-10 &  31.8\%\\ 
              \midrule   
              Median all  cases &   23 & 8.6e-02   &   271  &  7.0e-08 & 31.6\% \\ 
             \bottomrule
\end{tabular}
\label{Unclassified}
\end{table}
In the following, we compute infeasibility certificates and the volume of the remaining unclassified input space for a range of test cases. For this purpose, we run Algorithm~\ref{ALG_IF} with the number of iterations $N_1$ set to 1000. We evaluate the remaining estimated volume for the bound tightening $V^{BT}$ according to \eqref{eq:V_BT} and for the separating hyperplanes described as a convex polytope by running the volume approximation algorithm in \cite{emiris2018practical}. In Table~\ref{Unclassified}, the dimensionality $|x|$, the reduced unclassified volume $V^{BT}$ after bound tightening, the number of hyperplanes $|HP|$, and the reduced unclassified volume $V^{HP}$ enclosed by the separating hyperplanes is listed. Note that both volumes are defined with respect to the unit hypercube $x \in [0\,1]^{|x|}$ normalized by the original power system limits with volume $1$, i.e., $10^0$. 

We make several observations. First, the bound tightening results in a moderate reduction in input dimensionality of several orders of magnitude ($10^{-1}$~to~$10^{-4}$) for most test cases. Second, the infeasibility certificates based on hyperplanes enable further substantial reductions in the unclassified volume. As a result, the total unclassified volume compared to the unit hypercube is reduced between 2 and 40 orders of magnitude ($10^{-2}$~to~$10^{-40}$). The median of the unclassified volume is $10^{-8}$. This means that in order to identify one sample inside the unclassified volume, we would have to uniformly draw $10^8$ samples from the original bounds on the input $x$. This highlights the necessity of first computing the infeasibility certificates to be able to identify the secure space. The number of hyperplanes is below 1000 for most test cases, indicating that Algorithm~\ref{ALG_IF} has obtained a good estimation of the unclassified volume. For the four test cases for which 1000 hyperplanes are added, the unclassified input volume could be further reduced by increasing $N_1$.

To allow for comparability between test cases \blue{with different number of degrees of freedom}, we propose to use a metric defined as $\tfrac{-\log_{10}(V)}{|x|}$. The metric is motivated as follows: If one wants to sample 10 steps in each dimension, i.e., $10^{|x|}$, then this metric quantifies by how much the exponent is reduced. Note that the value obtained in percent is not the dimensionality reduction itself but relates to the reduction in the orders of magnitudes of the dimensionality. This value is between 14.8\% and 72.2\% for all test cases, showcasing the general applicability of the proposed infeasibility certificate for AC-OPF problems.

\subsection{Dataset Creation for PGLib-OPF Networks}
We create datasets of operating points classified based on their feasibility with respect to AC-OPF problems including N-1 security and uncertainty. To this end, we first draw a number of samples $N_2 = 10^4$ from the inside of the remaining unclassified volume described in Table~\ref{Unclassified} and obtain a detailed security boundary description following the approach in Section~\ref{sec:boundary}. We fit a multivariate normal distribution with $s_{\text{red}}=0.25$ and classify $N_3 = 10^5$ samples as secure or insecure following the approach in Section~\ref{sec:inside}. \begin{table}
\caption{Created Datasets for AC-OPF Problems} \begin{tabular}{l c c c c}
\toprule 
Power system case &  Boundary & Inside  & Overall & Overall \\
 &   & (MVND) & secure & points \\
 &  $N_2 = 10^4 $& $N_3 = 10^5$ &  & \\
   \midrule
                              \multicolumn{5}{c}{AC-OPF without N-1 security and without uncertainty} \\
                              \midrule
               \textit{case3\_lmbd}  &  69.5\%   &   36.5\%  &    40.6\%   &    114'389 \\ 
                  \midrule
               \textit{case5\_pjm}  &  68.6\%   &   69.4\%  &    69.3\%   &    125'432 \\ 
                  \midrule
              \textit{case14\_ieee}  &  73.3\%   &   59.0\%  &    61.0\%   &    147'047 \\ 
                 \midrule
          \textit{case24\_ieee\_rts}  &  66.8\%   &   44.3\%  &    48.7\%   &    131'158 \\ 
             \midrule
              \textit{case30\_ieee}  &  75.0\%   &   50.2\%  &    54.0\%   &    124'944 \\ 
                 \midrule
              \textit{case39\_epri}  &  57.2\%   &   29.9\%  &    33.9\%   &    154'635 \\ 
                \midrule
              \textit{case57\_ieee}  &  58.9\%   &   35.2\%  &    38.9\%   &    150'865 \\ 
                 \midrule
          \textit{case73\_ieee\_rts}  &  63.9\%   &   51.1\%  &    52.7\%   &    222'730 \\ 
             \midrule
             \textit{case118\_ieee}  &  53.2\%   &   47.0\%  &    47.6\%   &    209'996 \\ 
                \midrule
         \textit{case162\_ieee\_dtc}  &  50.0\%   &   40.1\%  &    41.7\%   &    129'165 \\ 
            \midrule
             \textit{case200\_tamu}  &  50.2\%   &   36.6\%  &    38.1\%   &    177'023 \\ 
                \midrule
             \textit{case300\_ieee}  &  50.0\%   &   32.6\%  &    34.7\%   &    163'087 \\ 
                \midrule
             \textit{case500\_tamu}  &  50.0\%   &   35.4\%  &    37.1\%   &    174'774 \\ 
              \midrule
              \multicolumn{5}{c}{AC-OPF considering N-1 security and uncertainty} \\
              \midrule        
              \textit{case39\_epri}  &  58.2\%   &   78.2\%  &    75.2\%   &    139'756 \\  
                      \midrule        
               \textit{case162\_ieee\_dtc}   &  50.0\%   &   17.9\%  &    23.2\%   &    121'358  \\ 
               \midrule 
              Average all cases & 59.7\%  &  44.2\%   &   46.5\%    &  152'424 \\ 
            %
\bottomrule
\end{tabular}
    \vspace{-0.25cm}
\label{Table_Datasets}
\end{table}
In Table~\ref{Table_Datasets}, we list the characteristics of the obtained datasets. First, note that in the boundary identification stage, if the percentage of secure points is above 50\%, then sampling directly from the remaining unclassified volume results in identifying secure operating points. This is the case for the majority of test cases, demonstrating that the infeasibility certificate is able to provide a tight approximation of the secure spaces of non-convex AC-OPF problems. For the test cases where the sampling did not find any secure samples, the number of iterations for the feasibility certificate could be enlarged or other relaxations such as moment-based relaxations described in \cite{molzahn2018fnt} could be used to further reduce the unclassified space in Algorithm~\ref{ALG_IF}. Second, the results show that sampling from a multivariate normal distribution fitted to the boundary samples results in identification of a large number of secure samples. The resulting datasets are well balanced with on average 46.5\% secure samples. Note that this is an important metric for the successful application of data-driven methods such as neural networks \cite{wehenkel2012automatic}. The number of overall points is dependent on the number of \blue{additional} feasible samples identified by enforcing the generators' reactive power limits in the AC power flows and differs between the test cases.

Regarding the computational tractability, all simulations were carried out on a laptop and the dataset creation for the largest test cases took a few hours, with the most computationally intense task being the AC-OPF evaluations in the boundary identification and the optimization-based bound tightening \cite{sundar2018optimization}. By using high-performance computing and parallelizing both the boundary identification and the AC power flow computations, we expect that our approach can scale to systems with thousands of buses. \blue{The number of samples chosen for each stage of the dataset creation method needs to be adjusted for the data-driven application at hand, and depends among other factors on the problem dimensionality, the chosen classifier, and the desired prediction accuracy. A common approach is to train and evaluate the performance of a data-driven method on datasets of different sizes.}

\subsection{Training Neural Network Classifiers}
As an illustrative data-driven application, we evaluate the performance of a neural network classifier trained on several of the created datasets. The neural network predicts a binary classification, i.e., whether the input $x$ belongs to the class ``secure'' or ``insecure''. We choose neural network structures with five hidden layers where the numbers of neurons of each hidden layer selected to be 10 times the input dimension $|x|$. We split the dataset into a training set consisting of 85\% of all samples and a test set of the remaining 15\%. Note that the classifier has no information of the test set during training, and its performance is evaluated on the test set only. This gives a metric for how well the classifier generalizes to unseen data. We train the neural networks using TensorFlow \cite{tensorflow2015-whitepaper} with standard training parameters and 250 epochs.

\begin{table}[]
    \centering
        \caption{Test set accuracy of neural network classifiers}
    \begin{tabular}{l | c | c }
    \toprule %
     Power system case  & Full dataset & Only boundary \\
               \midrule
         \multicolumn{3}{c}{AC-OPF without N-1 security and without uncertainty} \\
         \midrule
       \textit{case14\_ieee} & 78.2\%  & 60.5\% \\
        \textit{case39\_epri} & 74.6\% & 38.5\% \\ 
        \textit{case162\_ieee\_dtc} & 84.4\% & 49.8\%\\
                  \midrule
         \multicolumn{3}{c}{AC-OPF including N-1 security and uncertainty} \\
         \midrule
                 \textit{case39\_epri} & 81.0\% & 80.4\% \\ 
        \textit{case162\_ieee\_dtc} & 93.4\% & 31.9\%\\
         \bottomrule
    \end{tabular}
    \label{ClassifierPerf}
    \vspace{-0.25cm}
\end{table}
Table~\ref{ClassifierPerf} shows the test set accuracy, i.e., the share of correctly predicted labels for the test set. First, we use 85\% of the full dataset for training and 15\% of the full dataset for testing. Second, we only use the boundary samples from Section~\ref{sec:boundary} as training data and then test on 15\% of the full dataset. This gives us an estimation of the benefit of the additional sampling from the fitted multivariate distribution in Section~\ref{sec:inside}. We observe that the neural network classifier is able to generalize from the training to the test set and achieve high accuracy when using the full dataset. To further increase the classification accuracy, deeper neural networks or a deep autoencoder to identify lower-dimensional features could be used. We observe that only relying on the boundary samples for prediction is not sufficient for most test cases, higlighting the importance of obtaining a representative dataset.

\section{Conclusion} \label{Sec:Con}
Successful application of data-driven methods in power systems requires datasets of sufficient size, covering a wide range of operating points. Creating a dataset that characterizes the security boundary and sufficiently covers both secure and insecure operating regions is a highly computationally demanding task, even for medium-sized systems, as we showed in \cite{Thams2019}. In this paper, we propose an efficient method to create such datasets. We focus on AC-OPF feasibility and N-1 security, as any operating point should first satisfy static security criteria. Future work will extend this to include dynamic security criteria, similar to \cite{Thams2019}. We develop an infeasibility certificate based on separating hyperplanes which is able to classify large portions of the input space as insecure. We show that the infeasibility certificates reduce the unclassified input space volumes significantly, by up to $10^{-40}$ compared to an initial normalized input space volume of $1$ (i.e., $10^0$) based on defined control variable bounds. Although the secure operating region is a very small portion of the original input space, our method is able to produce balanced datasets of secure and insecure operating points, a property desired for successful applications of data-driven methods. As an illustrative application, we used the generated datasets to assess the performance of neural network classifiers. Future work is directed towards (i)~utilizing convex restrictions from \cite{Lee2019,cui2019solvability} to characterize secure spaces and (ii)~exploiting high-performance computing.

\section*{Acknowledgements}
 The authors would like to thank the two reviewers for their insightful comments. 

\bibliographystyle{IEEEtran}

\IEEEtriggeratref{18}
\bibliography{Bib}

\end{document}